\documentclass[runningheads]{llncs}
\usepackage{graphicx}

\begin{document}
\title{Mobile Phone Usage Data for Credit Scoring}
%

\author{Henri Ots\inst{1}\and
Innar Liiv\inst{1}\and
Diana Tur\inst{2}}
\authorrunning{H. Ots et al.}
%
\institute{Tallinn University of Technolgoy, Akadeemia tee 15a, 12618 Tallinn , Estonia
\email{henriots@online.ee,innar.liiv@taltech.ee}
\and
Warsaw School of Economics, aleja Niepodległości 162, 02-554 Warszawa, Poland
\email{tur.diana@gmail.com}}
\maketitle              
\begin{abstract}
The aim of this study is to demostrate that mobile phone usage data can be used to make predictions and find the best classification method for credit scoring even if the dataset is small (2,503 customers).  
We use different classification algorithms to split customers into paying and non-paying ones using mobile data, and then compare the predicted results with actual results. There are several related works publicly accessible in which mobile data has been used for credit scoring, but they are all based on a large dataset. Small companies are unable to use datasets as large as those used by these related papers, therefore these studies are of little use for them. In this paper we try to argue that there is value in mobile phone usage data for credit scoring even if the dataset is small. 
We found that with a dataset that consists of mobile data based only on 2,503 customers, we can predict credit risk. The best classification method gave us the result 0.62 AUC (area under the curve). 

\end{abstract}

\section{Introduction}

Credit scoring helps in increasing the speed and consistency of the loan application processes and allows lending firms to automate their lending processes \cite{ref_10}. In this case, credit scoring significantly reduces human involvement in credit evaluation and lessens the cost of delivering credit \cite{ref_11}. Moreover, by using credit scores, financial institutions are able to quantify risks associated with granting credit to a particular applicant in a shorter period of time. According to Leonard \cite{ref_12}, a study done by a Canadian bank found that the time it took to process a consumer loan application was shortened from nine days to three after credit scoring was used. As such, the optimisation of the loan processing time means that time saved on processing could be utilised to address more complex aspects in the firm. Banaslak and Kiely \cite{ref_13} concluded that with the help of credit scores, financial institutions are able to make faster, better and higherquality decisions.

There are more than 2 billion people in world who do not have a bank account \cite{ref_1}. This makes it difficult to perform a credit evaluation exercise for these individuals. With the rise of big data and especially new kinds of novel data sources, various data alternatives can be used to explain the financial inclusion of these unbanked individuals. For instance, mobile usage data is a novel data source that can be employed successfully. Mobile phone usage data can been considered as good alternative data for credit scoring. 

To what extent can one tell your personality by simply looking at how you use your mobile phone? The use of standard carrier logs to determine the personality of a mobile phone user is a hot topic, which has generated tremendous interest. The number of mobile phone users has reached 6 billion worldwide and \cite{ref_2} service providers are allowing increasing access to phone logs to researchers and \cite{ref_6} commercial partners \cite{ref_7}. If predicted accurately, mobile phone datasets could provide a valuable and cost-effective method of surveying personalities. For example, marketers and phone manufacturing companies might seek to access dispositional information about their customers so as to design customised offers and promotions \cite{ref_28}. The human-computer interface field uses personality. Thus, it benefits from the appraisal of user dispositions using automatically collected data. Lastly, the ability to extract personality and other psychosocial variables from a large population might lead to unparalleled discoveries in the field of social sciences \cite{ref_29}.

The use of mobile phones to predict people’s personalities is a result of advancement in data collection, machine learning and computational social science which has made it possible to infer various psychological states and traits based on how people use their cell phones daily. For example, some studies have shown that people’s personality can be predicted based on the pattern of how they use social media such as Facebook or Twitter 
\cite{ref_30}, \cite{ref_31}, \cite{ref_32}. Other researchers have used information about people’s usage of various mobile applications such as YouTube, Internet Calendar, games and so on to make conclusions about their mood and personality traits \cite{ref_33}, \cite{ref_34}, \cite{ref_35}, \cite{ref_36}, \cite{ref_37}. While these approaches are remarkable, they require access to a wide-ranging information about a person’s entire social network. These limitations greatly weaken the use of such classification methods for large-scale investigations \cite{ref_38}. 

The main contribution of this paper is to demonstrate that mobile phone usage data, instead of classical loan application data collection and other semi-manual processes, can be successfully used to predict credit risk even if the dataset is small (2,503 customers). Such contribution is novel and necessary for academic discussions as well as practical applications in relevant industries. From practical engineering point of view, the final selection of important features (Table \ref{variables}) can be considered an equally strong contribution besides the main results. 

The general outline  of the paper is as follows: first, related work in this area is introduced (Section 2), secondly, experimental set-up is explained regarding data, measures, and experiment design (Section 3), followed by experimental results (Section 4) and conclusions (Section 5).

\section{Related Work}

Credit scoring can be best explained as the use of statistical models in the transformation of relevant data into numerical measures, which inform organisations assessing the credit trustworthiness of clients.  Essentially, credit scoring is simply an industrialisation or proxy of trust; a logical and further development of the subjective credit ratings first provided by 19th century credit bureaus \cite{ref_8}.

Numerous literature review focus on the development, application and evaluation of predictive models used in the credit sector \cite{ref_27}.  

These models determine the creditworthiness of an applicant based on a set of descriptive variables. Corporate risk models use data from a statement on financial position, financial ratios or macro-economic pointers, while retail risk models use data captured in the application form such as the customer’s transaction history \cite{ref_18}. The difference between variables used in corporate and retail models indicates that more challenges arise in consumer than corporate credit scoring. This paper focuses on the retail business.  

There are several indications about the use of mobile phone usage data in credit scoring corporate world. However, only very few relevant papers \cite{ref_related1}, \cite{ref_related2}, \cite{ref_related3}, \cite{ref_39}, \cite{ref_40}, \cite{ref_41} are open to wider research community. 

Björkegren and Grissen \cite{ref_39} use behavioural signatures in mobile phone data to predict default with an accuracy almost similar to that of credit scoring methods that use financial history. The approach was validated using call records matched to loan results for a sample of borrowers in a Caribbean country. Applicants in the highest quartile of risk according to the authors’ measure were six times more likely to default in payment than those in the lowest quartile. They used two different algorithms, Random Forest and Logistic regression. The result obtained with the Random Forest algorithm was 0.710 AUC (area under the curve) and with Logistic regression 0.760 AUC. The dataset included information on 7,068 customers from a South-American country \cite{ref_39}. 
Jose San Pedro et al. developed MobiScore \cite{ref_40}, a methodology used to build a model of the user’s financial risk using data collected from mobile usage. MobiScore \cite{ref_40} was using data on 60,000 real people obtained from telecommunication companies and financial service providers in a Latin American country. They used gradient boosting, support vector machine and linear regression models to solve the problem. AUC results with different combinations were between 64.1 and 72.5 \cite{ref_40}. 
Speakman et al. demonstrated \cite{ref_41} how to use boosted decision trees to create a credit score for underbanked populations, enabling them to access a credit facility that was previously denied due to the unavailability of financial data. Their research result was a 55\% reduction in default rates while simultaneously offering credit opportunities to a million customers that were given a 0 credit limit in the bank’s original model. The dataset contained 295,926 labelled examples with over 30 categorical and real-valued features. AUC results with the boosted decision trees algorithm were 0.764 and with logistic regression 0.74 \cite{ref_41}. 

\section{Experimental set-up}

\subsection{Data}

The dataset comprises of information on 2,503 customers who have obtained a consumer loan, and allows one to understand their previous payment behaviour. Any means of identification have been entirely removed from the data and consequently anything personal has been remove. Information was initially obtained with the consent of the customers. The dataset was collected from an anonymous European consumer lending company whose customers uses their mobile application to submit digital loan applications. 

Using their payment behaviour we are able to separate the trustworthy customers from the untrustworthy ones. Our target variable identifies untrustworthy customers as those, who have got a 90-day delay in payment of their instalments. Additionally, the dataset will include about 1,516 trustworthy customers without debts that exceed the 90-day limit. Conclusively, this will result in the percentages of the trustworthy and untrustworthy customers being 60.57\% and 39.43\%, respectively.

Android phone users can be requested to yield the following data about their device (see Table \ref{raw_data}). For this research we did not use phone numbers, calendar body texts or text messages (SMS). 

From among all the varying parameters, 22 variables were selected to be used in the experiments necessary for the research (the variables are shown in Table \ref{variables}). The variables were chosen by using manual review and statistical analysis of dependencies. We chose variables that were less dependent on each other. Using these variables, one of them is a categorical variable while others are numerical. In some experiments we discretized some numerical variables into bins so that their data type changed to categorical. 

\begin{table}
\caption{Raw data from Android phones.}\label{raw_data}
\begin{tabular}{|l|l|}
\hline
Data group &  Data description\\
\hline
Device & Device ID\\
Device & OS (operating system) version\\
Device & SDK (software development kit) version\\
Device & Release version\\
Device & Device \\
Device & Model\\
Device & Product\\
Device & Brand\\
Device & Display\\
Device & Hardware\\
Device & Manufacturer\\
Device & Serial\\
Device & User\\
Device & Host\\
Network & Network ID\\
Network & Carrier\\
Network & Operator\\
Network & Subscriber\\
Calendar & Calendar ID\\
Calendar & Title\\
Calendar & Date\\
Calendar & Body\\
Call info & Caller ID\\
Call info & Receiver (contact/unknown)\\
Call info & Type (incoming/outgoing/missed/unanswered)\\
Call info & Number\\
Call info & Date\\
Call info & Duration\\
Contact info & Contact ID\\
Contact info & Contact number\\
Installed apps & App ID\\
Installed apps & Package name\\
Installed apps & Label\\
Installed apps & Version name\\
Installed apps & Version code\\
Installed apps & Install date\\
SMS info & SMS ID\\
SMS info & Type (incoming/outgoing)\\
SMS info & Conversation\\
SMS info & Number\\
SMS info & Message lenght\\
SMS info & SMS date\\
SMS info & SMS ID\\
Images & Image ID\\
Images & Image date\\
Images & Image location\\
Data storage & Data storage ID\\
Data storage & Path\\
Data storage & Last modifed\\
\hline
\end{tabular}
\end{table}

\begin{table}
\caption{ Variables for experiments.}\label{variables}
\begin{tabular}{|l|l|l|}
\hline
Data group & Calculated data points & Data type\\
\hline
Call info & Average number of calls per month. & Numerical\\
Call info & Average number of incoming calls per month. & Numerical\\
Call info & Average number of outgoing calls per month. & Numerical\\
Call info & Average number of missed calls per month. & Numerical\\
Call info & Average number of unanswered calls per month. & Numerical\\
Call info & Average call duration. & Numerical\\
Call info & Average outgoing call duration. & Numerical\\
Call info & Average incoming call duration. & Numerical\\
Call info & Maximum outgoing call duration. & Numerical\\
Call info & Maximum incoming call duration. & Numerical\\
Images & Average number of images per month. & Numerical\\
Images & Average number of images made in distinct places per month. & Numerical\\
SMS info & Average number of SMSs per month. & Numerical\\
SMS info & Average number of incoming SMSs per month. & Numerical\\
SMS info & Average number of incoming SMSs per month from contacts. & Numerical\\
SMS info & Average number of incoming SMSs per month & Numerical\\
~ & from an unknown number. & ~ \\
SMS info & Average number of outgoing SMSs per month. & Numerical\\
SMS info & Average number of outgoing SMSs per month from a contact. & Numerical\\
SMS info & Average number of outgoing SMSs per month & Numerical\\
~ & from an unknown number. & ~ \\
SMS info & Average number of SMS conversations per month. & Numerical\\
Contacts & Number of contacts. & Numerical\\
Device & SDK version. & Categorical\\
\hline
\end{tabular}
\end{table}

\subsection{Measures}

Harris \cite{ref_64} notes that in the process of developing and reporting the credit scoring models it is pragmatic to differentiate between the training and the reporting phase. This is due to the need of the person to provide clarity on the type of the metric that was initially applied in the selection of model parameters. When denoting the metric adopted, it would be sensible to use the term evaluation metric in the training process. On the other hand, to report the model performance during the performance phase, the term performance metric will be adopted \cite{ref_64}.

In this analysis, both the performance metric and the primary model evaluation metric are represented by the region under the ROC (Receiver Operating Characteristic) curve called AUC. The ROC curve, often adopted by the AUC, illustrates a two-component aspect of differential performance where the sensitivity (\ref{sensitivity}) (i.e. the relative amount of the actual positives which is forecasted as positive) and the specificity (\ref{specificity}) (i.e. the proportion of actual negatives that are forecasted as being negative) are plotted on the Y and X axis, respectively. Normally, the AUC figure is demonstrated as in (\ref{AUC}) the figure below where S1 illustrates the total sum of the customer’s creditworthiness rank. In this, a score of 100\% shows that the person classifying can impeccably differentiate between the classes, and a score of fifty percentage shows a classifier possessing a minor quality of differentiation \cite{ref_65}.

\begin{equation}\label{sensitivity}
Sensitivity =  \frac{true\,positive}{\displaystyle true\,positive+false\,negative}
\end{equation}

\begin{equation}\label{specificity}
Specificity =  \frac{true\,negative}{\displaystyle true\,positive+true\,negative}
\end{equation}

\begin{equation}\label{AUC}
AUC =  \frac{(S1-sensitivity)+[(sensitivity+1)+0.5]}{\displaystyle sensitivity+specificity}
\end{equation}

\begin{equation}\label{testaccuracy}
Test\,accuracy =  \frac{true\,positive}{\displaystyle true\,positive+false\,positive}
                 + \frac{true\,negative}{\displaystyle false\,negative+true\,negative}
\end{equation}

Different metrics can also be applied and used to produce the working of the categories used herein. For instance, to check for the correctness (\ref{testaccuracy}) it can also be taken to be the measure of how correct those applying for credit on a held back data test are classified.
Several performances are often applied when reporting the performance of the classifier developed in this analysis. For instance, the test accuracy below has also been reported to be a measure of how precise the applicants of credit is. Subsequently, the slanted datasets are familiar, similar to what is happening with actual world credit, which scores the datasets making it irrelevant \cite{ref_65}.

\subsection{Experiment design}

We carried out four experiments with five different classification methods and considered AUC to be the performance parameter. As the author’s previous experiences have illustrated, there are no specific rules for working with alternative data. Accordingly, we carried out four experiments based on different pre-processing techniques. 

In the first experiment we included all the calculated variables. The SDK variable, which is categorical, needs to be encoded. The SDK data has to be converted into numbers to make them comparable. The SDK version comprises six different values (19, 20, 21, 22, 23, 24, 25), for which we generated dummy variables. The values of these parameters are either 1 or 0. As a result, there can be no missing information in the dataset. The second step in the data pre-processing is to scale all variables to make them comparable with each other. 

In the second experiment we used the same pre-processing techniques as in the first experiment, but we added backward elimination. The principle of Occam’s razor states that the \cite{ref_66} model needs to be as simple as possible until it achieves an acceptable level of performance on training data. This will help to avoid over-fitting the model. With backward elimination we can throw out variables with p-value (probability value) $>$0.05 and the highest p value. After that we can calculate a new combination of p values and continue the same process until we have a set of variables, all with p lower than 0.05.

In the third experiment we used the same pre-processing method as before but modified the variables. We used the optimal binning technique to group the variables. Optimal binning is a method of pre-processing categorical predictors where we set values for variables by grouping them into optimal bins. Its purpose is to reduce the impact of statistical noise \cite{ref_67}. 

In order to choose the classifier methods for the experimental part we used three parameters:  

•	How have they functioned in previous credit scoring research? 

•	How have they functioned in previous credit scoring research using mobile data? 

•	How have they functioned in the author’s practical work in credit scoring models? 

According to these three parameters we chose for our experiments the following methods: logistic regression, decision tree, random forest, SVM and neural networks. 

When organising benchmarks in pattern recognition, there is often the problem of determining the size of the test set that would give statistically significant results. The commonly adopted ratio is 8:2 according to the Pareto principle. 

According to research by Isabelle Guyon and the formula she found we can determine the example test size. The fraction of patterns reserved for the validation set should be inversely proportional to the square root of the number of free adjustable parameters. The ratio of the validation set (v) to the training set (t) is v/t, and the scales are ln(N/h-max), where N is the number of families of recognizers, and h-max is the largest complexity of those families. Each family of recognizers is characterised by its complexity, which may or may not be related to the VC-dimension (Vapnik–Chervonenkis dimension), the description length, the number of adjustable parameters, or other measures of complexity \cite{ref_68}.  
According to a small sample size of customers we chose three different test size examples for this research. The test sizes we chose were 10\%, 25\% and 40\%.  

Testing any combination of variables first results in all variables. We then chose only the variables with p$<$0.05 and finally binned the variables with p$<$0.05. The intervals of the variables can be determined in a variety of ways. For example, by using prior knowledge on the data. The boundaries of the intervals are usually defined beforehand.

\subsection{Experiments}
The experiments described in this chapter were done using the Python programming language and the Spyder environment. We also used numpy, matplotlib, panda and scikit-learn Python libraries for statistical analyses. 

Tables \ref{all_variables},  \ref{variables_less_p} and  \ref{variables_bin_less_p} show a representation of the performance of classification methods using mobile data. The results in Table \ref{all_variables} show the classifiers’ performances with all variables. The results in Table \ref{variables_less_p} show the classifiers’ performance with only the variables whose p value is lower than 0.05. Table  \ref{variables_bin_less_p} shows binned variables whose p value is lower than 0.05. The tables suggest the models created for the prediction of creditworthiness as illustrated by AUC on the suppressed datasets. To determine the importance of the variation in performance between the models we can take AUC as the main parameter to see which model had the best performance. Tables \ref{all_variables}, \ref{variables_less_p} and \ref{variables_bin_less_p} can also be compared for training accuracy, test accuracy and training time(s).  

The target variable chosen was 0 for a performing customer and 1 for a non-performing customer. A non-performing customer in this research is set as one who is 90 or more days overdue in paying their debt. According to Barisitz, \cite{ref_69} the rule of being 90 days overdue is most common in the European country from which the data for this research were collected.

\begin{table}
\caption{Showing comparative classifier performances with all variables.}\label{all_variables}

\begin{tabular}{|l|l|l|l|l|l|}
\hline
Classifier & Test size & Training accuracy & Test accuracy & AUC & Training\\

~ & ~ & ~ & ~ & ~ & time (s)\\

\hline
Logistic regression & Test size=0.10 & 0.62 & 0.62 & 0.51 & 0.005\\
Logistic regression & Test size=0.25 & 0.61 & 0.62 & 0.55 & 0.005\\
Logistic regression & Test size=0.40 & 0.63 & 0.61 & 0.56 & 0.001\\
Decision tree & Test size=0.10 & 1.00 & 0.57 & 0.54 & 0.093\\
Decision tree & Test size=0.25 & 1.00 & 0.56 & 0.54 & 0.074\\
Decision tree & Test size=0.40 & 1.00 & 0.56 & 0.54 & 0.052\\
Random forest & Test size=0.10 & 0.98 & 0.63 & 0.62 & 0.103\\
Random forest & Test size=0.25 & 0.98 & 0.60 & 0.52 & 0.076\\
Random forest & Test size=0.40 & 0.98 & 0.61 & 0.58 & 0.059\\
SVM & Test size=0.10 & 0.61 & 0.65 & 0.56 & 3.330\\
SVM & Test size=0.25 & 0.60 & 0.59 & 0.56 & 2.150\\
SVM & Test size=0.40 & 0.60 & 0.59 & 0.57 & 1.220\\
Neural networks & Test size=0.10 & 0.69 & 0.60 & 0.59 & 100.630\\
Neural networks & Test size=0.25 & 0.67 & 0.59 & 0.57 & 69.790\\
Neural networks & Test size=0.40 & 0.69 & 0.61 & 0.55 & 51.710\\
\hline
\end{tabular}
\end{table}

\begin{table}
\caption{Showing comparative classifier performances with variables were p-value is lower than 0.05.}\label{variables_less_p}
\begin{tabular}{|l|l|l|l|l|l|}
\hline
Classifier & Test size & Training accuracy & Test accuracy & AUC & Training\\
~ & ~ & ~ & ~ & ~ & time (s)\\
\hline
Logistic regression & Test size=0.10 & 0.68 & 0.62 & 0.54 & 0.004\\
Logistic regression & Test size=0.25 & 0.76 & 0.56 & 0.53 & 0.003\\
Logistic regression & Test size=0.40 & 0.77 & 0.55 & 0.50 & 0.003\\
Decision tree & Test size=0.10 & 1.00 & 0.56 & 0.53 & 0.032\\
Decision tree & Test size=0.25 & 1.00 & 0.55 & 0.53 & 0.029\\
Decision tree & Test size=0.40 & 1.00 & 0.58 & 0.55 & 0.023\\
Random forest & Test size=0.10 & 0.98 & 0.66 & 0.56 & 0.064\\
Random forest & Test size=0.25 & 0.98 & 0.62 & 0.59 & 0.056\\
Random forest & Test size=0.40 & 0.98 & 0.60 & 0.58 & 0.045\\
SVM & Test size=0.10 & 0.60 & 0.65 & 0.53 & 1.305\\
SVM & Test size=0.25 & 0.60 & 0.59 & 0.56 & 1.094\\
SVM & Test size=0.40 & 0.60 & 0.59 & 0.57 & 0.678\\
Neural networks & Test size=0.10 & 0.64 & 0.61 & 0.57 & 84.360\\
Neural networks & Test size=0.25 & 0.63 & 0.60 & 0.55 & 89.720\\
Neural networks & Test size=0.40 & 0.65 & 0.60 & 0.58 & 62.620\\
\hline
\end{tabular}
\end{table}

\begin{table}
\caption{Showing comparative classifier performances with variables are binned and p-value is lower than 0.05.}\label{variables_bin_less_p}
\begin{tabular}{|l|l|l|l|l|l|}
\hline
Classifier & Test size & Training accuracy & Test accuracy & AUC & Training\\
~ & ~ & ~ & ~ & ~ & time (s)\\
\hline
Logistic regression & Test size=0.10 & 0.68 & 0.62 & 0.54 & 0.004\\
Logistic regression & Test size=0.25 & 0.76 & 0.56 & 0.53 & 0.003\\
Logistic regression & Test size=0.40 & 0.77 & 0.55 & 0.50 & 0.003\\
Decision tree & Test size=0.10 & 0.74 & 0.59 & 0.53 & 0.060\\
Decision tree & Test size=0.25 & 0.76 & 0.56 & 0.53 & 0.003\\
Decision tree & Test size=0.40 & 0.78 & 0.55 & 0.50 & 0.003\\
Random forest & Test size=0.10 & 0.73 & 0.58 & 0.54 & 0.220\\
Random forest & Test size=0.25 & 0.75 & 0.53 & 0.52 & 0.022\\
Random forest & Test size=0.40 & 0.76 & 0.54 & 0.50 & 0.021\\
SVM & Test size=0.10 & 0.60 & 0.65 & 0.49 & 0.007\\
SVM & Test size=0.25 & 0.59 & 0.59 & 0.48 & 0.238\\
SVM & Test size=0.40 & 0.60 & 0.59 & 0.48 & 0.168\\
Neural networks & Test size=0.10 & 0.61 & 0.63 & 0.51 & 84.400\\
Neural networks & Test size=0.25 & 0.61 & 0.62 & 0.54 & 63.330\\
Neural networks & Test size=0.40 & 0.62 & 0.58 & 0.45 & 53.820\\
\hline
\end{tabular}
\end{table}

\section{Experimental results}

The empirical results consist of the performance estimates of five classifiers with three different combinations. The tables on the previous page report the AUCs of all five classifiers with all three experiment combinations.

Random forest provides the best average AUC level across experiments with different test sizes. Random forest also ranks the best AUC at 0.62 with all variables and a test size of 10. The second-best method was neural networks with the highest AUC and all variables using 10 for the test size.  

According to the author’s previous know-how as regards choosing a test size for a small dataset of 2,503 customers, we were able to take the most stable results at a 40 test size. With the test size being 40, we gained the best result from the first experiment with the random forest algorithm AUC=0.58, and the same result from the neural networks algorithm in the second experiment with only the variables where p$<$0.05. 

The weakest result overall was obtained from the SVM algorithm, which yielded very poor results in the second and third experiment, where AUC was below 0.50. The decision tree algorithm shows the most stable results across experiments and test sizes, having AUC between 0.50 and 0.55. 

Comparing the results with related works in Table \ref{related_works}, it is apparent that in this research, the AUC results are lower than in others. There is high correlation between the test size and AUC, and seeing how our sample size is only 2,503 customers compared to 7,068, 60,000 and 295,926 we can consider our results to be good.

\begin{table}
\caption{Comparing related works.}\label{related_works}
\begin{tabular}{|p{5cm} |l|p{4.3cm}|l|}
\hline
Work & Test set size & Method & AUC\\
\hline
 Behavior Revealed in Mobile Phone Usage Predicts Loan Repayment”, authors: Björkegren and Grissen, 2017 & 7,068 & Random Forest & 0.710\\
 \hline
Behavior Revealed in Mobile Phone Usage Predicts Loan Repayment”, authors: Björkegren and Grissen, 2017 & 7,068 & Logistic regression & 0.760\\
\hline
Mobile phone-based Credit Scoring”, authors: Skyler Speakman, Eric Mibuari,  Isaac Markus, Felix Kwizera, 
2017 & 60,000 & Linear regression & 0.725\\
\hline
MobiScore: Towards Universal Credit Scoring from Mobile Phone Data”, authors: Jose San Pedro,  Davide Proserpio and Nuria Oliver, 2015 & 295,926 & Boosted decision trees algorithm & 0.740\\
 \hline
MobiScore: Towards Universal Credit Scoring  from Mobile Phone Data”, authors: Jose San Pedro,  Davide Proserpio and Nuria Oliver, 2015 & 295,926 & Logistic regression & 0.760\\
  \hline
 Current research & 2,503 & Logistic regression & 0.540\\
  \hline
 Current research & 2,503 & Decision tree & 0.550\\
   \hline
 Current research & 2,503 & Random forest & 0.620\\
   \hline
 Current research & 2,503 & SVM & 0.570\\
   \hline
 Current research & 2,503 & Neural networks & 0.590\\

\hline
\end{tabular}
\end{table}

This study has two important theoretical contributions. First, based on the use of mobile data for credit scoring research, we can see that all the tested methods with all variables yielded a better result than in a random study.  

Secondly, we empirically demonstrate that the best method for credit scoring based on mobile data is the random forests classification method with AUC 0.62. 

Our research on mobile data scoring will make it possible for other organizations in the financial sector to use mobile data for their credit scoring. While prior three researches on this subject showed that mobile data is only useful with big datasets, we maintain that it can yield positive results even with a small dataset. Thus, this knowledge can now be used in small or medium-sized companies as well. 

\section{Conclusions}

For the past three decades, financial risk forecasting has been one of the main areas of growth in statistics and probability modelling. People often think of the term ‘financial risk’ in relation with portfolio management when it comes to pricing of options among other financial instruments. The main challenge for consumer loan firms over the past years has been reaching the huge sector of unbanked customers. There are more than 2 billion people in world who do not have a bank account \cite{ref_1} and the number of mobile phone users has reached 6 billion worldwide \cite{ref_2}. Few conceptual works have been posited with a research subject that brings together credit scoring and mobile data. 

This paper is based on a synthesis of earlier academic research with new experiments and argues that mobile phone usage data can give positive results for credit scoring even with a small dataset. Our findings also reveal that the best model in terms of mobile data usage for credit scoring is the decision tree method. 

If finance companies want to have more accurate data on those customers who are more likely to pay back their loans, they need to find alternative data sources such as mobile phone data. This will a give huge advantage to finance companies in third world countries where most people do not have any bank history – the only data they have is their mobile phone data.  

We hope this study opens up further discussion and advances theory to generate a more accurate understanding of how we can use mobile data to make predictions and added value. This paper could spark discussion not only for financial sector companies but also in the field of insurance or fraud prevention, where mobile data can help make predictions. 

There are many ways in which future studies could elaborate on this subject. One way is to look at algorithms in more depth and try to come up with more accurate models. Making predictions on mobile data can be used in other sectors as well, not only in finance. It is very probable that if we can predict customers’ payment behaviour based on mobile data, we could also predict their insurance or fraud risk. There are multiple research possibilities in the field of alternative data sources such as mobile data that could add value for businesses. In the modern world we have many technical solutions at our disposal that create and gather data every day.  

\bibliographystyle{splncs04}
\bibliography{references}

\end{document}